# A General Theory of Piping Transportation: Unifying System Dynamics for Resilience and Sustainable Development

## Samuel Darwisman

Institut Transportasi dan Logistik Trisakti, Jakarta, Indonesia
*Corresponding author. Email:* [23c606001088@student.itltrisakti.ac.id](mailto:23c606001088@student.itltrisakti.ac.id)

**Abstract**

The science of pipeline transport is currently governed by a collection of fragmented, discipline-specific theories that are inadequate for addressing the systemic challenges of 21st-century infrastructure. This paper introduces and formalizes a new, unified theory: the General Theory of Piping Transportation (GTPT), formulated by Darwisman. The GTPT posits that a pipeline system is a complex socio-technical entity whose state and long-term viability are determined by the fully coupled interaction of three interdependent domains: Physical Dynamics ($\Phi$), Life-Cycle Dynamics ($\Lambda$), and Socio-Economic Dynamics ($\Sigma$). This paper presents the core postulates of the GTPT, which are derived from a systematic synthesis of the fragmented existing literature. The prescriptive power of the theory is illustrated by contrasting the strategic outcomes derived from the GTPT against those from classical theories. By defining resilience as the primary design objective and operationalizing the UN Sustainable Development Goals (SDGs), the GTPT provides a new theoretical foundation for the design, management, and governance of infrastructure across all critical sectors.

## 1. Introduction: The Theoretical Deficit in Pipeline Science and the Imperative for a New Paradigm

### 1.1. The Central Role of Pipeline Infrastructure in the Global Economy

Pipeline systems are the arterial infrastructure of the global economy, the unseen yet indispensable conduits for modern industrial society (Dubey et al., 2019). They facilitate the large-scale, continuous transport of a vast array of substances that underpin nearly every aspect of economic activity. These include primary energy resources like oil and natural gas, transitional fuels such as liquefied natural gas (LNG), and future energy carriers like hydrogen and captured carbon dioxide ($CO_2$) (Raj et al., 2024; Liu et al., 2026). Beyond energy, pipelines are critical for transporting essential resources like fresh water, managing wastewater, and moving industrial feedstocks and processed materials in sectors ranging from mining to agriculture and manufacturing (Rennó &

Lemgruber, 2024). The efficiency and reliability of these networks are direct determinants of industrial competitiveness, energy security, and public welfare. Consequently, investment in pipeline infrastructure is a key enabler for achieving numerous Sustainable Development Goals (SDGs), from providing clean water and sanitation (SDG 6) and affordable, clean energy (SDG 7), to fostering industrial innovation and resilient infrastructure (SDG 9).

## 1.2. The Fragmentation of "Classical Piping Transportation Theory"

Despite their systemic importance, the body of knowledge guiding the design and management of these multi-billion-dollar assets—which we collectively term "Classical Piping Transportation Theory"—suffers from a fundamental and increasingly dangerous deficit: fragmentation. The field does not exist as a unified science but as a collection of siloed, discipline-specific sub-theories, each with its own models, assumptions, and optimization objectives. These include:

   a. **Fluid Dynamics and Thermodynamics:** This foundational stream focuses on the internal mechanics of flow. It employs principles like the Navier-Stokes equations, conservation laws, and empirical correlations for friction factors to model flow regimes, predict pressure drop, and optimize hydraulic efficiency (Moody, 1944; Stuckenbruck, 2024). Its primary objective is to maximize throughput while minimizing energy consumption.

   b. **Structural Mechanics and Material Science:** This discipline concentrates on the physical integrity of the pipeline as a structure. It develops sophisticated models to analyze stress, strain, and material degradation, predicting and preventing failures from phenomena such as flow-induced vibrations (FSI), fatigue, and corrosion (Nabudda et al., 2014; Wasim & Djukic, 2022). Its objective is to ensure asset integrity and safety within a defined operational lifespan.

   c. **Geotechnical Engineering:** This field analyzes the critical interaction between the pipeline and its surrounding environment, particularly for buried or subsea infrastructure. It models soil-structure interaction, the impact of geohazards, and seepage-induced erosion (El Shamy & Aydin, 2008; Wewer et al., 2021). Its objective is to ensure the stability of the asset within its physical setting.

   d. **Economic Optimization and Operations Research:** This managerial discipline focuses on the logistical and financial efficiency of the pipeline network. It employs mathematical programming and algorithms to optimize scheduling, routing, and operational costs, often with the primary objective of minimizing capital and operational expenditures (CAPEX/OPEX) (Sarkar & Arya, 2022; Arya et al., 2022).

## 1.3. The Consequence of Fragmentation: Brittle Designs and a "Resilience Gap"

While each of these theoretical streams is invaluable, their siloed nature is their greatest weakness. This fragmentation promotes a paradigm of localized optimization, where technical efficiency is pursued often at the expense of systemic resilience. This creates "brittle" infrastructure—assets optimized for a narrow set of predictable conditions but fragile in the face of systemic shocks.

Prior systematic reviews by the author have consistently illuminated this fragmentation. Research on integrating SDGs into logistics has revealed that human and organizational factors are often disconnected from technological strategies (Darwisman et al., 2025). Analyses of green port strategies underscore a persistent "policy-practice gap," where sustainability goals fail due to a lack of integration with economic and governance realities (Darwisman, 2025a; Darwisman et al., 2025b). Foundational work on reconstructing transportation cost theory further cemented the obsolescence of classical linear models in the face of modern non-linear, stepwise cost structures (Darwisman, 2025c; Liu, 2025). This collective evidence points to an urgent need for a new theoretical foundation. This paper aims to fill this void by proposing and formalizing the General Theory of Piping Transportation (GTPT).

## 2. Methods: Theoretical Construction Methodology

The General Theory of Piping Transportation (GTPT) was constructed through a rigorous, multi-stage theoretical synthesis methodology designed to build a new, comprehensive framework from a critique of existing knowledge.

### 2.1. Phase 1: Systematic Deconstruction and Critique of Existing Theories

The foundation of this work was a comprehensive critique of the fragmented literature, drawing upon the author's extensive prior work in systematic reviews (Darwisman, 2025a, 2025b, 2025c; Darwisman et al., 2025). This involved a structured deconstruction of the core assumptions, boundaries, and optimization objectives of the key sub-disciplines. The primary finding was the systemic failure of these theories to account for the coupled, non-linear interactions between technical, environmental, and socio-economic variables.

### 2.2. Phase 2: Axiomatic Formulation (Darwisman's Synthesis 1)

Based on the identified theoretical gaps, a deductive process was used to formulate a new set of foundational principles, or postulates. This synthesis, formulated by Darwisman, represents the first core theoretical contribution of this paper. These postulates are logical propositions designed to create a more robust theoretical structure, explicitly redefining the system's boundaries, primary objective, and core dynamics.

### 2.3. Phase 3: Conceptual Synthesis and Framework Modeling

The new postulates were then operationalized into a cohesive conceptual framework, structuring the theory into three core domains—Physical Dynamics ($\Phi$), Life-Cycle Dynamics ($\Lambda$), and Socio-

Economic Dynamics (Σ). This process is visually represented in the conceptual models (Figures 1, 2, and 3) and qualitatively defined in the descriptive tables (Tables 1, 2, and 3).

## 3. Results: The General Theory of Piping Transportation (GTPT) – Postulates and Framework

The primary result of this research is the formal articulation of the GTPT itself, defined by its core postulates and conceptual framework.

### 3.1. The Core Postulates of the GTPT

**Postulate 1: The Principle of Socio-Technical Holism.** *A pipeline system is not merely a physical object but a complex, adaptive socio-technical system. Its state cannot be described by physical variables alone but must be defined by a state vector that encompasses its physical, life-cycle, and socio-economic dimensions.*

**Postulate 2: The Primacy of Resilience.** *For long-life infrastructure operating under uncertainty, the primary objective of design and management is not the optimization of short-term efficiency but the maximization of long-term systemic resilience.*

**Postulate 3: The Law of Coupled Dynamics.** *The domains of Physical Dynamics (Φ), Life-Cycle Dynamics (Λ), and Socio-Economic Dynamics (Σ) are non-separable and fully coupled. Any perturbation in one domain will propagate non-linearly through the others.*

### 3.2. The Conceptual Framework and Darwisman's Synthesis Statements

The GTPT framework operationalizes these postulates. The following sections detail each domain and present the novel theoretical syntheses that form the core of this work.

**Table 1. A Comparative Analysis of Classical Theory vs. the General Theory of Piping Transportation (GTPT).**

| Dimension | Classical Theory Paradigm | GTPT Paradigm (Darwisman's Formulation) |
|---|---|---|
| **Core Objective** | Maximize Efficiency | Maximize Systemic Resilience |
| **System Boundary** | The physical asset | The asset, its value chain, and its societal/environmental context |
| **Analytical Approach** | Siloed, linear optimization | Integrated, multi-domain, non-linear co-optimization |
| **Failure Definition** | Technical malfunction | Any event compromising long-term systemic function and value |

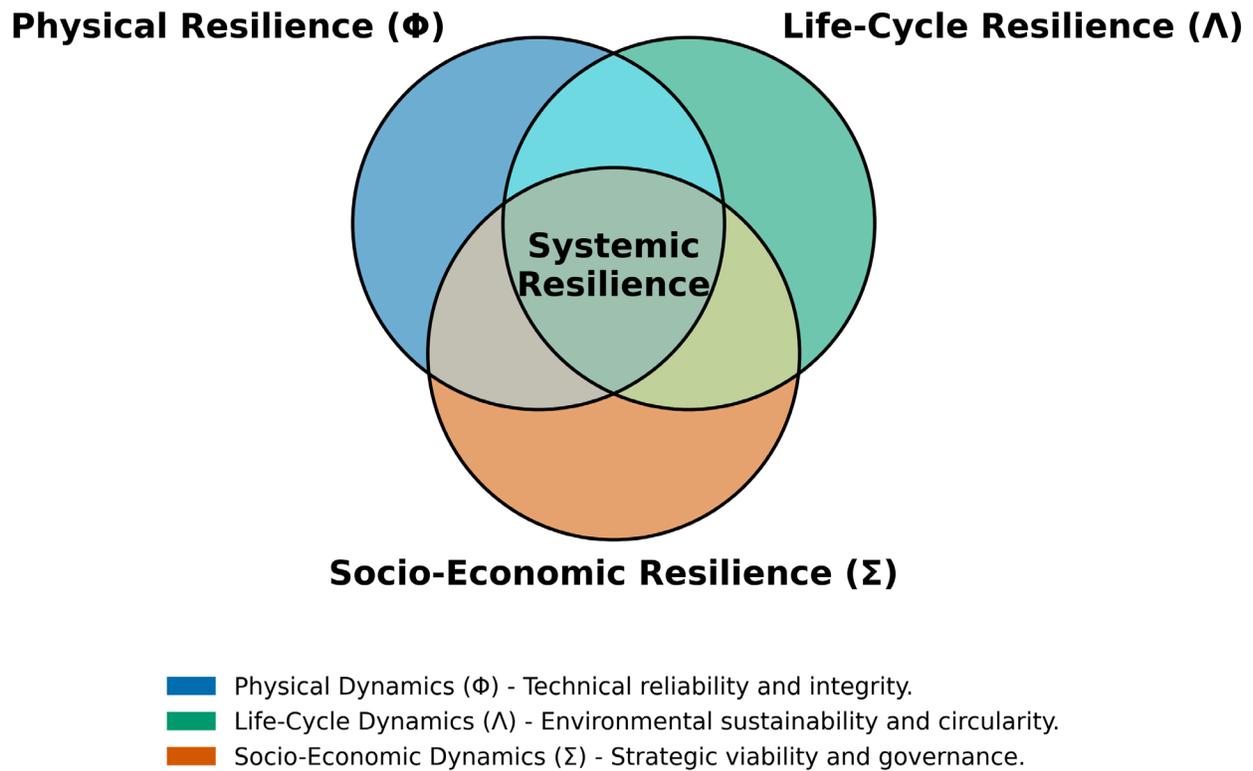

**Figure 1. The three core, interdependent domains of resilience in the General Theory of Piping Transportation (GTPT).**

Systemic resilience is an emergent property of the coupled interaction between these domains.

**Domain I: Physical Dynamics (Φ) and the Theory of Resilience-Based Integrity**

> **Darwisman's Synthesis 2:** The Theory of Resilience-Based Integrity. Synthesizing the literature on multi-phase flow instabilities (Gong et al., 2020), pressure pulsations (Zheng et al., 2024), and transient events (Nabudda et al., 2014), Darwisman proposes that physical design should not be optimized for a single, static load, but for "dynamic robustness" across a spectrum of probable and improbable stresses.

**Domain II: Life-Cycle Dynamics (Λ) and the Principle of Endogenous Externalities**

> **Darwisman's Synthesis 3:** The Principle of Endogenous Externalities. Drawing from literature on Life Cycle Assessment (LCA) (Chohan et al., 2023; Karkhanis et al., 2025) and environmental impact assessment (Padash & Ataee, 2019), Darwisman posits that environmental and social impacts are not "externalities" but are endogenous variables with a quantifiable "shadow price" that must be included directly within the primary optimization function.

**Domain III: Socio-Economic Dynamics (Σ) and the Theory of Socio-Economic Viability**

> **Darwisman's Synthesis 4:** The Theory of Socio-Economic Viability. Synthesizing insights from risk management (Derse et al., 2022), policy alignment (Vasilyev et al., 2019), and strategic management (Zhang et al., 2024), Darwisman proposes that "Social License to Operate" and "Regulatory Adaptability" are quantifiable assets whose value can be modeled as a reduction in the probability of catastrophic project delays, thereby impacting the risk-adjusted Net Present Value (RA-NPV).

**Table 2. The Three Core Domains of the GTPT Framework and Their Theoretical Novelty.**

| Domain | Classical Focus | GTPT Novelty (Darwisman's Synthesis) |
|---|---|---|
| Φ (Physical) | Performance Optimization | **Resilience-Based Integrity:** Design for a spectrum of dynamic stresses. |
| Λ (Life-Cycle) | Compliance with Regulations | **Endogenous Externalities:** Internalize environmental/social impacts as direct costs. |
| Σ (Socio-Economic) | Cost Minimization | **Socio-Economic Viability:** Quantify social license and regulatory adaptability as assets. |

**4. Discussion: The Unifying Power and Broad Applicability of the GTPT**

The GTPT's primary contribution is its unifying power, providing a common theoretical language for analyzing disparate challenges across numerous industrial sectors.

**4.1. The GTPT in Practice: A New Decision-Making Framework**

The theory fundamentally changes how strategic decisions are evaluated. Figure 2 illustrates this conceptual shift. A classical approach seeks a "local optimum" along a single dimension. The GTPT compels a search for a "global optimum" within a multi-dimensional "resilience space."

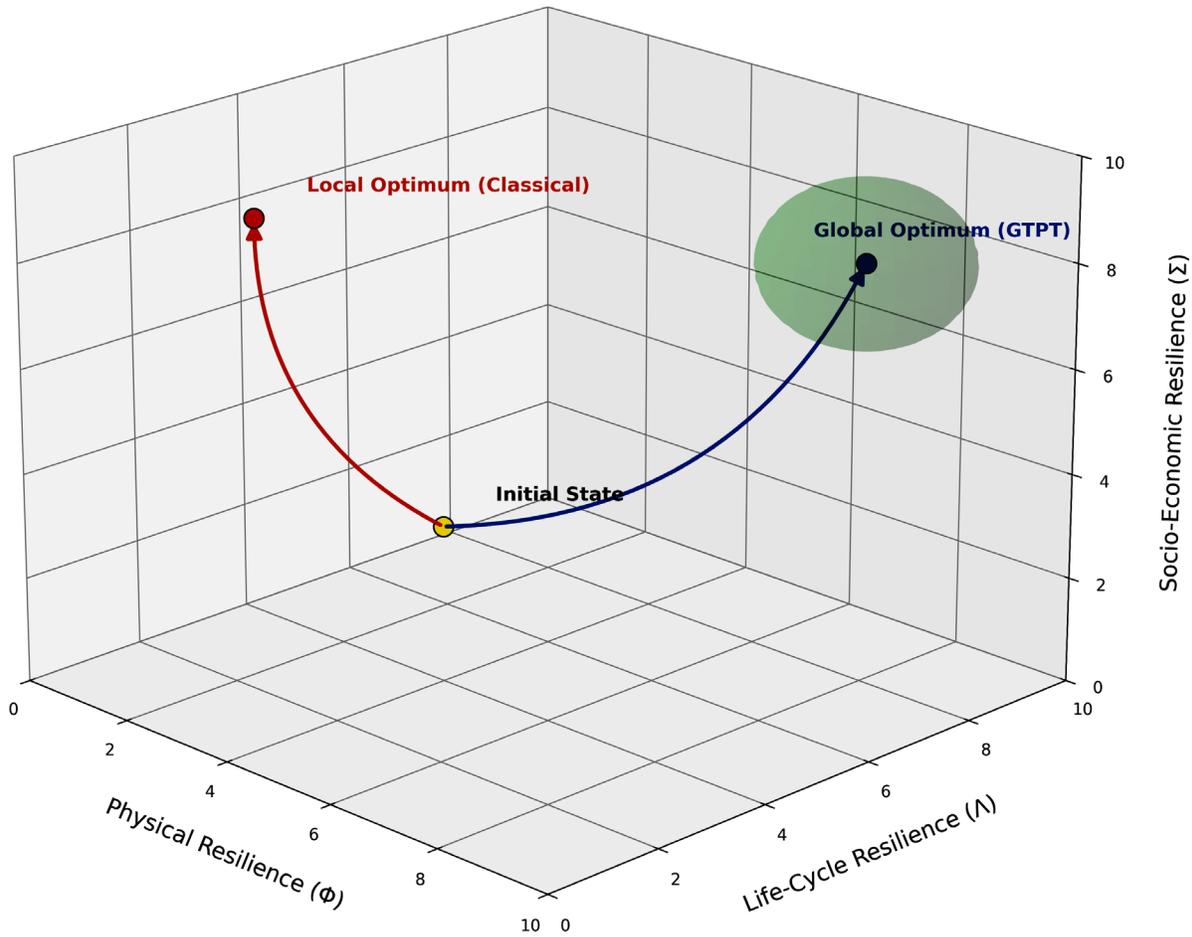

**Figure 2. Conceptual Framework for Design Decision-Making. The classical approach (red path) seeks a local optimum of cost efficiency.**

The GTPT approach (blue path), guided by Darwisman's syntheses, navigates a multi-dimensional resilience space to find a globally optimal, resilient solution.

### 4.2. Broad Applicability of the Theory across Industrial Sectors

The true test of a "general theory" is its ability to provide explanatory and prescriptive power across a wide range of applications.

a. **Energy Sector (Oil, Gas, Hydrogen, $CO_2$):** The GTPT provides a critical framework for managing the energy transition. For the hydrogen economy, it allows a holistic comparison between repurposing existing pipelines versus building new ones, balancing the physical risks of hydrogen embrittlement ($\Phi$) (Raj et al., 2024), the life-cycle costs of materials ($\Lambda$) (Chohan et al., 2023), and the public safety and regulatory challenges ($\Sigma$).
b. **Water and Wastewater Management:** The GTPT reframes the problem of aging infrastructure from "fixing leaks" to "managing systemic resilience," justifying investment in smart water grids by quantifying the coupled benefits in physical reliability (Φ), resource conservation (Λ), and public health (Σ).
c. **Circular Economy and Industrial Waste Management:** The GTPT is the first theory to explicitly frame pipelines as enablers of industrial symbiosis. It provides the business case for investing in "circular pipelines" by co-optimizing: the technical reliability for handling complex slurries (Φ) (Lan et al., 2013); the enormous environmental benefits of diverting waste from landfills (Λ); and the creation of new, resilient circular value chains (Σ).
d. **Maritime Logistics and Port Operations (Shipping):** At ports, the GTPT guides the strategic design of "resilient port infrastructure" for new green fuels, integrating the physical requirements for cryogenic handling (Φ), the life-cycle spill risks (Λ), and the complex web of international maritime regulations and economic competition (Σ).
e. **Construction, District Energy, and Urban Development:** For district energy networks, the GTPT justifies investment in advanced insulation by balancing higher CAPEX (Σ) against long-term physical performance (reduced heat loss, Φ) and massive life-cycle energy savings (Λ). In construction, it provides a framework for selecting pipe materials based on their life-cycle impact (Karkhanis et al., 2025) and resilience to future urban stresses.

**Table 3. Illustrative Application of the GTPT Across Different Industrial Sectors.**

| Sector | Physical Challenge (Φ) | Life-Cycle Challenge (Λ) | Socio-Economic Challenge (Σ) |
|---|---|---|---|
| **Hydrogen Energy** | Material Embrittlement, High Pressure | Energy Intensity of Production | Public Safety Perception, Regulation |
| **Waste Management** | Abrasive/Corrosive Flow, Blockage | Leachate Contamination Risk | "NIMBY" Opposition, Permitting |
| **Urban Water** | Aging Infrastructure, Leakage | Pumping Energy Consumption | Public Health, Water Scarcity |

**4.3. The GTPT Application Cycle: A Framework for Strategic Implementation**

To ensure the theory is not just descriptive but also prescriptive, Darwisman proposes the GTPT Application Cycle (Figure 3). This iterative process shows how organizations can operationalize the theory for continuous improvement and strategic adaptation, making it a universally applicable management tool.

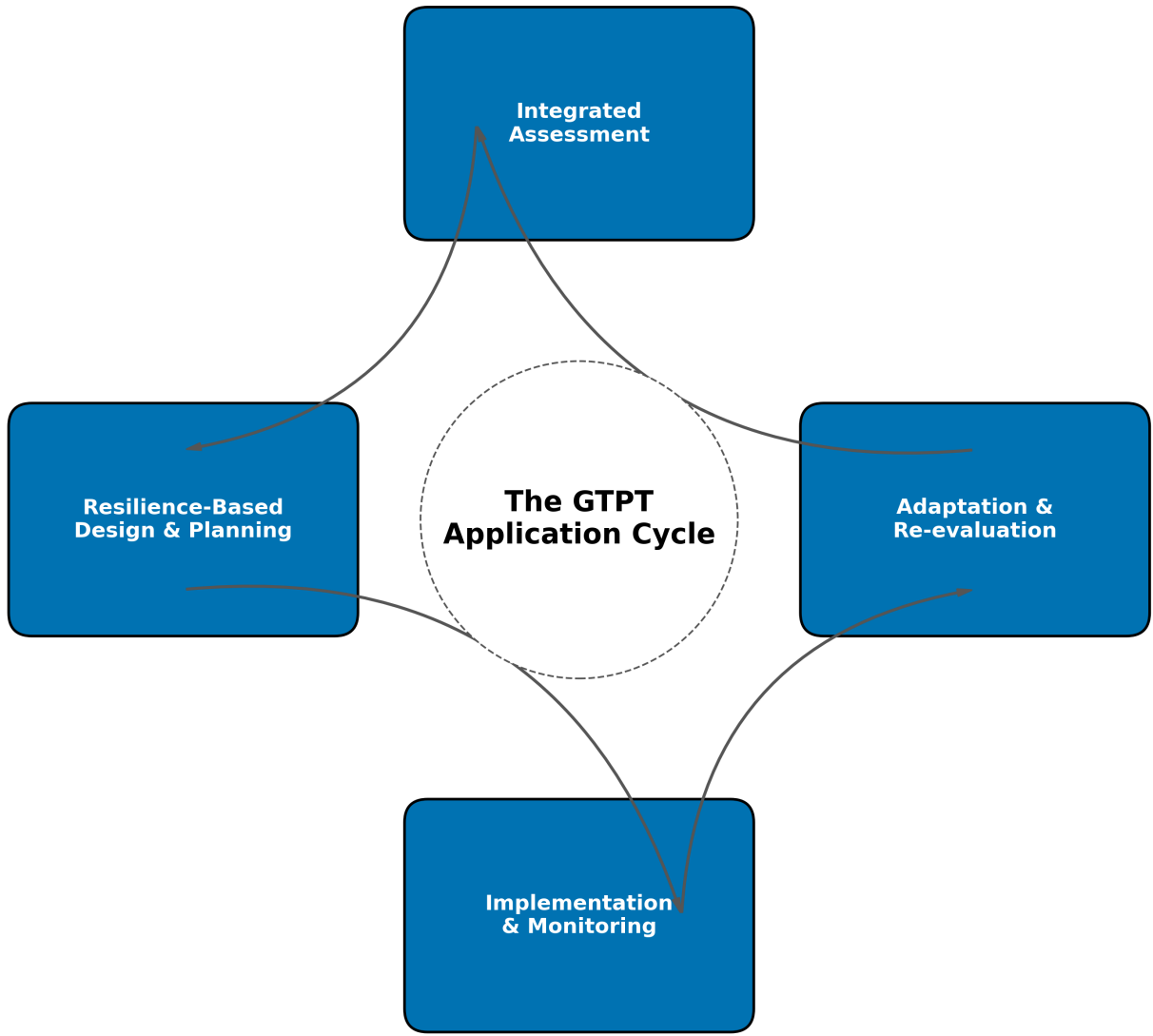

**Figure 3. The GTPT Application Cycle.**

This universal framework illustrates how the theory can be operationalized as an iterative strategic management process for designing and managing resilient pipeline systems in any sector or country.

The cycle consists of four key stages:

1. **Integrated Assessment:** This initial stage involves a holistic data gathering and analysis of the system's context across all three GTPT domains (Φ, Λ, Σ).

2. **Resilience-Based Design & Planning:** In this stage, multi-objective optimization is performed, guided by the GTPT's postulates, to identify a design that represents the global optimum of resilience.

3. **Implementation & Monitoring:** The chosen design is constructed and operated with a robust monitoring system to collect real-time data on performance across all three domains.

4. **Adaptation & Re-evaluation:** The monitored data feeds back into the assessment stage, allowing for continuous learning and adaptation, ensuring the system can evolve and maintain its resilience over time.

## 4.4 Limitations and a Call for Future Research

As a new theoretical proposition, the GTPT has limitations that also serve as a roadmap for future research.

a. **Limitation 1: Conceptual Nature.** The GTPT is a conceptual framework. Future research must focus on developing computational models and software tools that can operationalize it.

b. **Limitation 2: Context-Dependency of Metrics.** The weighting of variables is context-dependent. Future empirical case studies are needed to develop a typology of resilience functions for different sectors.

c. **Limitation 3: Data Intensity.** A full GTPT analysis requires vast amounts of data. Future research is needed on developing robust methodologies for quantifying qualitative variables, leveraging big data and AI.

## 5. Conclusion

This paper has introduced the **General Theory of Piping Transportation (GTPT)**, formulated by **Darwisman**. By establishing the postulates of Socio-Technical Holism, the Primacy of Resilience, and the Law of Coupled Dynamics, and supported by novel theoretical syntheses, the GTPT provides a necessary evolution beyond the fragmented paradigms of classical theory. It offers a unified framework that redefines the core objective of pipeline science and management, shifting the focus from isolated efficiency to integrated, systemic resilience. The theory's broad applicability demonstrates its power as a truly "general" theory, offering a robust foundation for academia, industry, and policymakers to guide the strategic design and governance of infrastructure that is not only built to last, but built for a complex and changing world.

## 6. References

Anvari, M., Baldin, A., Clees, T., Klaassen, B., Nikitin, I., Nikitina, L., & Pott, S. (2024). Stability of dynamic fluid transport simulations. *Journal of Physics: Conference Series*, *2701*(1), 012009. https://doi.org/10.1088/1742-6596/2701/1/012009

Arya, A. K., Jain, R., Yadav, S., Bisht, S., & Gautam, S. (2022). Recent trends in gas pipeline optimization. *Materials Today: Proceedings*, *57*, 1455–1461. https://doi.org/10.1016/j.matpr.2021.11.232

Bales, P., Kolb, O., & Lang, J. (2009). Hierarchical modelling and model adaptivity for gas flow on networks. *Lecture Notes in Computer Science*, *5544*(PART 1), 337–346. https://doi.org/10.1007/978-3-642-01970-8_33

Blaic, S., Matko, D., & Geiger, G. (2015). The simulation of multi-batch pipelines by a multiscale method. *Proceedings - 8th EUROSIM Congress on Modelling and Simulation, EUROSIM 2013*, 460–465. https://doi.org/10.1109/EUROSIM.2013.84

Bouzit, F., Bouzit, M., & Mokeddem, M. (2022). Study of the Rheological Behaviour and the Curvature Radius Effects on a Non Newtonian Fluid Flow in a Curved Square Duct. *International Journal of Engineering Research in Africa*, *59*, 225–238. https://doi.org/10.4028/p-ll8x57

Cely, A., Hammer, M., Andersen, H., Yang, T., Nekså, P., & Wilhelmsen, Ø. (2022). *Thermodynamic Model Evaluations for Hydrogen Pipeline Transportation*. Society of Petroleum Engineers. https://doi.org/10.2118/209626-MS

Chandratilleke, T. T., & Nursubyakto, A. (2003). Numerical prediction of secondary flow and convective heat transfer in externally heated curved rectangular ducts. *International Journal of Thermal Sciences*, *42*(2), 187–198. https://doi.org/10.1016/S1290-0729(02)00018-2

Chen, W., Deng, G., Zhou, C., & Yuan, Z. (2016). Numerical simulation and experiment study on the resistance loss of glue pipeline transportation. *2016 17th International Conference on Electronic Packaging Technology, ICEPT 2016*, 138–141. https://doi.org/10.1109/ICEPT.2016.7583106

Chohan, I. M., Ahmad, A., Sallih, N., Bheel, N., Ali, M., & Deifalla, A. F. (2023). A review on life cycle assessment of different pipeline materials. *Results in Engineering*, *19*, 101325. https://doi.org/10.1016/j.rineng.2023.101325

Cividini, A., & Gioda, G. (2004). Finite-element approach to the erosion and transport of fine particles in granular soils. *International Journal of Geomechanics*, *4*(3), 191–198. https://doi.org/10.1061/(asce)1532-3641(2004)4:3(191)

Darwisman, S. (2025a). *Integrating Global Multimodal Transportation Sustainability: A Systematic Literature Review Aligning with SDG 9 and 13*. [Preprint]. ScienceOpen. https://doi.org/10.14293/PR2199.002206.v1

Darwisman, S. (2025b). *From Mine to Metropolis: A Systematic Review of Green Mining Innovations for Sustainable Urban Logistics*. [Unpublished manuscript]. Faculty of Management and Business, Institut Transportasi dan Logistik Trisakti. https://doi.org/10.38035/dijemss.v7i2.6033

Darwisman, S. (2025c). *Reconstructing Transportation Cost Planning Theory: A Multi-Layered Framework Integrating Stepwise Functions, AI-Driven Dynamic Pricing, and Sustainable Autonomy*. [Unpublished manuscript]. Faculty of Management and Business, Institut Transportasi dan Logistik Trisakti. https://doi.org/10.48550/arXiv.2512.10494

Darwisman, S., Jayawibawa, M., Bugiman, H., Asrori, A. F., Hermawan, A. D., & Abidin, Z. (2025). Green Port Strategies for Indonesian Maritime Logistics: A Systematic Review of Mining Barging and General Cargo. *Dinasti International Journal of Education Management and Social Science, 7*(2). https://doi.org/10.38035/dijemss.v7i2.5796

Darwisman, S., Simanjuntak, Y. R., Sihombing, S., & Setyawati, A. (2025). Integrating Sustainable Development Goals (SDGs) into Human Resource Management: A Systematic Review, Critical Appraisal, and Framework for the Logistics and Transportation Sector. *Jurnal Ekonomi Manajemen Sistem Informasi, 6*(6), 4351–4360. https://doi.org/10.38035/jemsi.v6i6

Darwisman, S., Sulistiyono, A., Karsa, A., Hendra, W., & Maemunah, S. (2025). Managing the Energy Transition for Sustainable Land Transportation in Indonesia: A Systematic Review of Policy and Socio-Economic Implications for Road and Rail. *Dinasti International Journal of Education Management and Social Science, 7*(2). https://doi.org/10.38035/dijemss.v7i2.5749

De Freitas, R. V. N., Sondermann, C. N., Patricio, R. A. C., Figueiredo, A. B., Bodstein, G. C. R., Rachid, F. B. F., & Baptista, R. M. (2018). Numerical study of two-phase flow in a horizontal pipeline using an unconditionally hyperbolic two-fluid model. *ASME International Mechanical Engineering Congress and Exposition, Proceedings (IMECE), 7*. https://doi.org/10.1115/IMECE201887571

Derse, O., Oturakci, M., & Dagsuyu, C. (2022). Risk Analysis Application to Hazardous Material Transportation Modes. *Transportation Research Record, 2676*(3), 586–597. https://doi.org/10.1177/03611981211052961

Dubey, P. N., Verma, R. K., Verma, G., & Reddy, G. R. (2019). Design and Analysis of Piping and Support. In *Textbook of Seismic Design: Structures, Piping Systems, and Components* (pp. 379–418). https://doi.org/10.1007/978-981-13-3176-3_11

Ebrahimi-Mamaghani, A., Sotudeh-Gharebagh, R., Zarghami, R., & Mostoufi, N. (2019). Dynamics of two-phase flow in vertical pipes. *Journal of Fluids and Structures, 87*, 150–173. https://doi.org/10.1016/j.jfluidstructs.2019.03.010

El Shamy, U., & Aydin, F. (2008). Multiscale modeling of flood-induced piping in river levees. *Journal of Geotechnical and Geoenvironmental Engineering*, *134*(9), 1385–1398. https://doi.org/10.1061/(ASCE)1090-0241(2008)134:9(1385)

Garbai, L., & Halász, G. (2022). Extending the Validity of Basic Equations for One-dimensional Flow in Tubes with Distributed Mass Sources and Varying Cross Sections. *Periodica Polytechnica Mechanical Engineering*, *66*(3), 237–243. https://doi.org/10.3311/PPme.20079

Gong, J., Shi, B., Chen, Y., & Song, S. (2020). Submarine multiphase pipeline transport containing natural gas hydrate and its plugging risk prevention and control. *Natural Gas Industry*, *40*(12), 133–142. https://doi.org/10.3787/j.issn.1000-0976.2020.12.015

Ibrahim, R. I., Odah, M. K., & Al-Mufti, A. (2019). An Overview on the Recent Techniques for Improving the Flowability of Crude Oil in Pipelines. *IOP Conference Series: Materials Science and Engineering*, *579*(1), 012054. https://doi.org/10.1088/1757-899X/579/1/012054

Il'ichev, A. T., Sumskoi, S. I., & Shargatov, V. A. (2018). Unsteady Flows in Deformable Pipes: The Energy Conservation Law. *Proceedings of the Steklov Institute of Mathematics*, *300*(1), 68–77. https://doi.org/10.1134/S0081543818010054

Jiménez-Cabas, J., Piñeres-Espitia, G., Rodelo-Serrano, E., Pamucar, D., Ramirez-Leon, H., Butt, S. A., Naz, S., & Simic, V. (2025). Modelling and control of a multi-product pipeline transportation process. *International Journal of Information Technology (Singapore)*, *17*(9), 5409–5419. https://doi.org/10.1007/s41870-025-02419-x

Karimov, K., Khudjaev, M., & Akhmedov, A. (2021). Modeling fluid outflow from a channel consisting of three different segments. *E3S Web of Conferences*, *258*, 08021. https://doi.org/10.1051/e3sconf/202125808021

Karkhanis, E., Kaushal, V., Thakre, G., & Najafi, M. (2025). SimaPro-Based Comparative Ecological Impact Assessment of CIPP Renewal and Open-Trench Pipeline Replacement. In *Pipelines 2025: Planning and Design - Proceedings of Sessions of the Pipelines 2025 Conference* (Vol. 1, pp. 257–267). https://doi.org/10.1061/9780784486368.028

Karpenko, M. (2024). AIRCRAFT HYDRAULIC DRIVE ENERGY LOSSES AND OPERATION DELAY ASSOCIATED WITH THE PIPELINE AND FITTING CONNECTIONS. *Aviation*, *28*(1), 1–8. https://doi.org/10.3846/aviation.2024.20946

Karpenko, M., & Bogdevičius, M. (2020). Investigation into the hydrodynamic processes of fitting connections for determining pressure losses of transport hydraulic drive. *Transport*, *35*(1), 108–120. https://doi.org/10.3846/transport.2020.12335

Khismatullin, A. S., & Turkin, D. A. (2025). Integration of Automated Management Systems for Enterprises' Transport and Technical Services. In *Lecture Notes in Electrical Engineering* (Vol. 1324 LNEE, pp. 65–74). https://doi.org/10.1007/978-3-031-82494-4_7

Khujaev, I., Shadmanova, G., Mamadaliev, K., & Aminov, K. (2020). *Mathematical modeling of transition processes due to a change in gas consumption at the ends of the inclined section of the gas*. 2020 International Conference on Information Science and Communications Technologies, ICISCT 2020. https://doi.org/10.1109/ICISCT50599.2020.9351523

Lan, G., Jiang, J., Li, D. D., Yi, W. S., Zhao, Z., & Nie, L. N. (2013). Research on numerical simulation and protection of transient process in long-distance slurry transportation pipelines. *IOP Conference Series: Materials Science and Engineering*, *52*(TOPIC 7), 072008. https://doi.org/10.1088/1757-899X/52/7/072008

Layeghi, M., & Seyf, H. R. (2008). Fluid flow in an annular microchannel subjected to uniform wall injections. *Journal of Fluids Engineering, Transactions of the ASME*, *130*(5). https://doi.org/10.1115/1.2911655

Liang, Y., Chen, J., & Chen, L. (2011). Mathematical model for piping erosion based on fluid-solid interaction and soils structure. *Geotechnical Special Publication*, (217 GSP), 109–116. https://doi.org/10.1061/47628(407)14

Liang, Y., He, G., Fang, L., Wu, M., Gao, J., Li, Y., & Li, F. (2017). Research advances in the influence of temperature on the sequential transportation in product pipeline. *Kexue Tongbao/Chinese Science Bulletin*, *62*(22), 2520–2533. https://doi.org/10.1360/N972016-00275

Liu, D., Wang, S., Liu, C., Wang, W., Chen, Z., Chen, S., Zhang, S., Du, S., & Hong, B. (2026). Advancements in CO2 pipeline transportation technology: a bibliometric analysis and knowledge mapping study. *Fuel*, *407*, 137498. https://doi.org/10.1016/j.fuel.2025.137498

Liu, J. (2025). A New Bounding Procedure for Transportation Problems with Stepwise Costs. *Mathematics*, *13*(22), 3709. https://doi.org/10.3390/math13223709

Liu, S., Zhang, Y., Yang, J., Sun, D., Wang, Y., Kang, H., & Zhang, S. (2025). Research status and development trends of hydrogen pipeline transportation based on knowledge mapping. *You Qi Chu Yun/Oil and Gas Storage and Transportation*, *44*(1), 10–25. https://doi.org/10.6047/j.issn.1000-8241.2025.01.002

Liu, Z., Guan, Z., Zhao, J., Wu, P., Zhang, T., & Qi, M. (2024). Analysis of research hotspots in oil & gas storage and transportation based on reports in Engineering Fronts and databases for retrieval. *You Qi Chu Yun/Oil and Gas Storage and Transportation*, *43*(11), 1201–1211. https://doi.org/10.6047/j.issn.1000-8241.2024.11.001

Liu, Z. E., Long, W., Chen, Z., Littlefield, J., Jing, L., Ren, B., El-Houjeiri, H. M., Qahtani, A. S., Jabbar, M. Y., & Masnadi, M. S. (2024). A novel optimization framework for natural gas transportation pipeline networks based on deep reinforcement learning. *Energy and AI*, *18*, 100434. https://doi.org/10.1016/j.egyai.2024.100434

Luan, Z., Zhong, L., Zhang, H., Lan, C., Yang, Y., Du, X., & Liu, Y. (2024). Kinetic energy correction coefficient for rectangular drainage channels. *Physics of Fluids*, *36*(3), 037141. https://doi.org/10.1063/5.0185941

Mäkká, K., Tušer, I., & Šiser, A. (2024). Assessing Risks Associated with the Transportation of Hazardous Materials: A Case Study of Transport Company. *Transport Means - Proceedings of the International Conference*, *2024-October*, 78–83. https://doi.org/10.5755/e01.2351-7034.2024.P78-83

Mayakova, S. A., & Privalov, L. Y. (2020). Thermal and hydraulic calculation in a model of industrial branched pipeline transporting a single-phase liquid flow. *E3S Web of Conferences*, *224*, 01048. https://doi.org/10.1051/e3sconf/202022401048

McGlinchey, D. (2023). Simulation of Pneumatic Conveying. In *Simulations in Bulk Solids Handling: Applications of DEM and other Methods* (pp. 145–176). https://doi.org/10.1002/9783527835935.ch5

Monette, C., & Pettigrew, M. J. (2004). Fluidelastic instability of flexible tubes subjected to two-phase internal flow. *Journal of Fluids and Structures*, *19*(7), 943–956. https://doi.org/10.1016/j.jfluidstructs.2004.06.003

Moody, L. F. (1944). Friction Factors for Pipe Flow. *Journal of Fluids Engineering, Transactions of the ASME*, *66*(8), 671–678. https://doi.org/10.1115/1.4018140

Morita. (2020). Discretization of the system of equations for fluid flow through porous media. *Developments in Petroleum Science*, *70*, 261–272. https://doi.org/10.1016/B978-0-12-823825-7.00007-7

Mostafa, A., Moustafa, K., & Elshaer, R. (2023). Impact of Fixed Cost Increase on the Optimization of Two-Stage Sustainable Supply Chain Networks. *Sustainability*, *15*(18), 13949. https://doi.org/10.3390/su151813949

Nabudda, K., Limtragool, J., & Piyasin, S. (2014). Study on failure of pipeline due to pressure surge effect. *Applied Mechanics and Materials*, *487*, 348–351. https://doi.org/10.4028/www.scientific.net/AMM.487.348

Neganov, D. A. (2019). Formation of knowledge base and data banks for justification of strength reliability of oil and oil products pipeline transportation system. *Science and Technologies: Oil and Oil Products Pipeline Transportation*, *9*(5), 488–504. https://doi.org/10.28999/2541-9595-2019-9-5-488-504

Padash, A., & Ataee, S. (2019). Prioritization of environmental sensitive spots in studies of environmental impact assessment to select the preferred option, based on AHP and GIS compound in the gas pipeline project. *Pollution*, *5*(3), 671–685. https://doi.org/10.22059/poll.2019.270349.546

Peng, D.-Y., & Robinson, D. B. (1976). A New Two-Constant Equation of State. *Industrial and Engineering Chemistry Fundamentals*, *15*(1), 59–64. https://doi.org/10.1021/i160057a011

Rahman, K., Parvin, S., & Khan, A. H. (2024). Analysis of two-phase flow in the porous medium through a rectangular curved duct. *Experimental and Computational Multiphase Flow*, *6*(1), 67–83. https://doi.org/10.1007/s42757-023-0159-9

Raj, A., Larsson, I. A. S., Ljung, A.-L., Forslund, T., Andersson, R., Sundström, J., & Lundström, T. S. (2024). Evaluating hydrogen gas transport in pipelines: Current state of numerical and experimental methodologies. *International Journal of Hydrogen Energy*, *67*, 136–149. https://doi.org/10.1016/j.ijhydene.2024.04.140

Rennó, M., & Lemgruber, N. (2024). Introduction: The Pipeline Life Cycle. In *Handbook of Pipeline Engineering* (pp. 3–33). https://doi.org/10.1007/978-3-031-33328-6_1

Rysbaiuly, B., & Aigul, S. (2020). Mathematical modeling of the heat transfer equation in the pipeline. *ACM International Conference Proceeding Series*, 3410802. https://doi.org/10.1145/3410352.3410802

Saad, J. (2023). Modeling of steady-state non-isothermal flow for natural gas transmission in the buried pipeline-case study. *AIP Conference Proceedings*, *2809*(1), 030014. https://doi.org/10.1063/5.0143701

Sarkar, A., & Arya, A. K. (2022). A Survey on Optimization Parameters and Techniques for Crude Oil Pipeline Transportation. *Smart Innovation, Systems and Technologies*, *292*, 561–574. https://doi.org/10.1007/978-981-19-0836-1_43

Seliverstov, S., & Kantemirov, I. (2019). State equations for oil & gas pipelines subjected to bending strain and other forces. *AIP Conference Proceedings*, *2167*, 020315. https://doi.org/10.1063/1.5132182

Shestakov, R. A., Rezanov, K. S., Matveeva, Y. S., & Vanchugov, I. M. (2022). IMPROVED MATHEMATICAL MODEL OF THE MAIN PIPELINE WITH LOOPING. *Bulletin of the Tomsk Polytechnic University, Geo Assets Engineering*, *333*(2), 123–131. https://doi.org/10.18799/24131830/2022/2/3325

Sheludchenko, B., Slusarenko, I., Pluzhnikov, O., Shubenko, V., Biletsky, V., & Borovskyi, V. (2021). Analytical criterion for the strength of bonded-dispersed gels during pipeline transportation. *Scientific Horizons*, *24*(2), 9–15. https://doi.org/10.48077/scihor.24(2).2021.9-15

Skempton, A. W., & Brogan, J. M. (1994). Experiments on piping in sandy gravels. *Geotechnique*, *44*(3), 449–460. https://doi.org/10.1680/geot.1994.44.3.449

Soave, G. (1972). Equilibrium constants from a modified Redlich-Kwong equation of state. *Chemical Engineering Science*, *27*(6), 1197–1203. https://doi.org/10.1016/0009-2509(72)80096-4

Stavropoulou, M., Papanastasiou, P., & Vardoulakis, I. (1998). Coupled wellbore erosion and stability analysis. *International Journal for Numerical and Analytical Methods in Geomechanics*, *22*(9), 749–769. https://doi.org/10.1002/(SICI)1096-9853(199809)22:9<749::AID-NAG944>3.0.CO;2-K

Sterpi, D. (2003). Effects of the erosion and transport of fine particles due to seepage flow. *International Journal of Geomechanics*, *3*(1), 111–122. https://doi.org/10.1061/(ASCE)1532-3641(2003)3:1(111)

Stuckenbruck, S. (2024). Flow Mechanics of Pipelines. In *Handbook of Pipeline Engineering* (pp. 73–121). https://doi.org/10.1007/978-3-031-33328-6_4

Sumskoi, S. I., Sofin, A. S., & Lisanov, M. V. (2016). Developing the model of non-stationary processes of motion and discharge of single- and two-phase medium at emergency releases from pipelines. *Journal of Physics: Conference Series*, *751*(1), 012025. https://doi.org/10.1088/1742-6596/751/1/012025

Tang, Y., & Yang, T. (2018). Post-buckling behavior and nonlinear vibration analysis of a fluid-conveying pipe composed of functionally graded material. *Composite Structures*, *185*, 393–400. https://doi.org/10.1016/j.compstruct.2017.11.032

Tang, Y., Wang, Y.-X., Chen, D.-Y., & Yang, T.-Z. (2025). Nonlinear dynamics and vibration suppression of fiber-reinforced composite three-dimensional pipes transporting liquid-gas two-phase flow with nonlinear energy sink. *Physics of Fluids*, *37*(8). https://doi.org/10.1063/5.0277914

Tmur, A. B. (2009). Identification of fluid flow velocity and pressure distributions in pipelines. *IFAC Proceedings Volumes (IFAC-PapersOnline)*, *13*(PART 1), 2167–2172. https://doi.org/10.3182/20090603-3-RU-2001.0535

Udoetok, E. S. (2019). Pressure drop in liquid-liquid core-annular fluid flow. *International Journal of Fluid Mechanics Research*, *46*(3), 229–238. https://doi.org/10.1615/InterJFluidMechRes.2018021520

Valle, J. D., Vera, S., Tovar, J., & Asuaje, M. (2025). ESTIMATION OF LOCALIZED ENERGY LOSS COEFFICIENTS IN A 90° ELBOW FOR CRUDE OIL AND WATER TWO-PHASE FLOW USING CFD. *American Society of Mechanical Engineers, Fluids*

*Engineering Division (Publication) FEDSM*, 2. https://doi.org/10.1115/FEDSM2025-158607

Van Beek, V. M., De Bruijn, H. T. J., Knoeff, J. G., Bezuijen, A., & Förster, U. (2010). Levee failure due to piping: A full-scale experiment. *Geotechnical Special Publication*, (210 GSP), 283–292. https://doi.org/10.1061/41147(392)27

Vasilyev, G. G., Leonovich, I. A., & Latifov, K. A. (2019). On the methodology of risk-oriented standardization of safety parameters at the design and construction of gas and oil pipelines. *Bezopasnost' Truda v Promyshlennosti*, *2019*(2), 84–90. https://doi.org/10.24000/0409-2961-2019-2-84-90

Wang, Y., et al. (2025). Theoretical analysis of gas–liquid–solid three-phase flow in air-lift pipelines for mineral particles. *Ocean Engineering*, *339*. https://doi.org/10.1016/j.oceaneng.2025.122091

Wang, Y.-X., Tang, Y., & Yang, T.-Z. (2024). Nonlinear mechanic analysis of a composite pipe conveying solid-liquid two-phase flow. *Applied Ocean Research*, *144*. https://doi.org/10.1016/j.apor.2024.103905

Wang, Y.-X., Tang, Y., & Yang, T.-Z. (2025). Nonlinear vortex-induced vibration analysis of a fiber-reinforced composite pipes transporting liquid-gas two-phase flow. *Communications in Nonlinear Science and Numerical Simulation*, *142*. https://doi.org/10.1016/j.cnsns.2024.108516

Wasim, M., & Djukic, M. B. (2022). External corrosion of oil and gas pipelines: a review of failure mechanisms and predictive preventions. *Journal of Natural Gas Science and Engineering*, *100*. https://doi.org/10.1016/j.jngse.2022.104467

Wei, L.-X., Wen, J.-B., Liu, Y., & Xu, Y.-C. (2011). *Process calculation method study and application of oil-gas multiphase transportation pipeline*. Asia-Pacific Power and Energy Engineering Conference, APPEEC. https://doi.org/10.1109/APPEEC.2011.5748554

Wewer, M., Aguilar-López, J. P., Kok, M., & Bogaard, T. (2021). A transient backward erosion piping model based on laminar flow transport equations. *Computers and Geotechnics*, *132*, 103992. https://doi.org/10.1016/j.compgeo.2020.103992

Woxenius, J. (2007). Generic framework for transport network designs: Applications and treatment in intermodal freight transport literature. *Transport Reviews*, *27*(6), 733–749. https://doi.org/10.1080/01441640701358796

Wu, X., Lu, H., Huang, K., Yuan, Z., & Sun, X. (2015). Mathematical model of leakage during pressure tests of oil and gas pipelines. *Journal of Pipeline Systems Engineering and Practice*, *6*(4), 04015001. https://doi.org/10.1061/(ASCE)PS.1949-1204.0000195

Wu, X., Zhao, H., Zhang, T., & Liu, X. (2025). Experimental Study on Transport of Fine Particles and Evolution of Acoustic Emission Characteristics in Porous Media Seepage Process. *Pure and Applied Geophysics*, *182*(10), 4283-4301. https://doi.org/10.1007/s00024-025-03793-0

Yagubov, E. Z., Tshadaya, N. D., & Yagubov, Z. X. (2013). Multichannel pipelines for transportation of oil and gas and revitalizing of worn-out oil and gas pipelines. *SOCAR Proceedings*, *2013*(1), 57–63. https://doi.org/10.5510/OGP20130100145

Zang, Z., Fan, W., & Hu, C. (2025). Numerical Investigation of Local Scour Below a Submarine Pipeline on Sand Wave Seabeds Under Current Conditions. *Water (Switzerland)*, *17*(22). https://doi.org/10.3390/w17223279

Zhang, B., et al. (2024). A mathematical framework for modeling and designing a long distance cryogenic liquefied natural gas pipeline: A practical study and analysis in Shandong, China. *Energy Conversion and Management*, *312*, 118516. https://doi.org/10.1016/j.enconman.2024.118516

Zhang, C., Han, Z., Ma, B., Yang, Z., Liu, Y., Hu, Y., Wang, Z., & Zhao, K. (2025). Coupling Changes in Pressure and Flow Velocity in Oil Pipelines Supported by Structures. *Processes*, *13*(9). https://doi.org/10.3390/pr13092932

Zhang, C., Sun, X., Li, Y., & Zhang, X. (2017). Numerical simulation of hydraulic characteristics of cyclical slit flow with moving boundary of tube-contained raw materials pipelines hydraulic transportation. *Nongye Gongcheng Xuebao/Transactions of the Chinese Society of Agricultural Engineering*, *33*(19), 76–85. https://doi.org/10.11975/j.issn.1002-6819.2017.19.010

Zhang, C., Sun, X., Li, Y., Zhang, X., Zhang, X., Yang, X., & Li, F. (2018). Effect of fluid-structure interaction on internal flow field characteristics of tube-contained raw material pipeline hydraulic transportation. *Nongye Gongcheng Xuebao/Transactions of the Chinese Society of Agricultural Engineering*, *34*(18), 299–307. https://doi.org/10.11975/j.issn.1002-6819.2018.18.037

Zhang, H., Tu, R., Zhang, X., Yang, X., Su, Y., Xia, Y., Yan, J., & Qiu, R. (2024). Efficiency evaluation of product oil pipeline under transportation market competition. *You Qi Chu Yun/Oil and Gas Storage and Transportation*, *43*(10), 1180–1188. https://doi.org/10.6047/j.issn.1000-8241.2024.10.011

Zhang, J., & Guo, L. (2024). Latest progress in hydrogen pipeline transportation technology. *Huagong Jinzhan/Chemical Industry and Engineering Progress*, *43*(12), 6692–6699. https://doi.org/10.16085/j.issn.1000-6613.2023-2164

Zheng, G., Xu, P., Li, L., & Fan, X. (2024). Investigations of the Formation Mechanism and Pressure Pulsation Characteristics of Pipeline Gas-Liquid Slug Flows. *Journal of Marine Science and Engineering*, *12*(4). https://doi.org/10.3390/jmse12040590